\documentclass[10pt,twocolumn,letterpaper]{article}

\usepackage[pagenumbers]{wacv}
\usepackage{booktabs}
\usepackage{multirow}
\usepackage{tabularx}
\usepackage{array}
\usepackage{makecell}
\ifdefined\pdfobjcompresslevel
\fi

\usepackage{amsmath}
\usepackage{amssymb}
\usepackage{booktabs}
\usepackage{array}
\usepackage{graphicx}
\usepackage{microtype}
\usepackage[T1]{fontenc}
\usepackage{adjustbox}

\newcommand{\method}{CPathOGen}
\newcommand{\cellvit}{CellViT++}
\newcommand{\hovernet}{HoVer-Net}
\newcommand{\stardist}{StarDist}

\definecolor{wacvblue}{rgb}{0.21,0.49,0.74}
\usepackage[pagebackref,breaklinks,colorlinks,allcolors=wacvblue]{hyperref}

\def\wacvPaperID{*****}
\def\confName{WACV}
\def\confYear{2027}

\title{CPathOGen: Spatially and Morphologically Controlled H\&E Counterfactuals for Probing Pathology Models}
\author{Samarth Singhal\\
University of North Dakota\\
{\tt\small samarth.singhal@und.edu}
\and
Varang Rai\\
Google\\
{\tt\small varangrai@google.com}}

\begin{document}
\maketitle

\begin{abstract}
Computational pathology models infer biologically and clinically meaningful outcomes from histology, but their predictions are shaped by complex, intertwined tissue signals whose roles are important to understand. Common pixel- and feature-space perturbations can produce implausible tissue, making model responses difficult to interpret. We introduce CPathOGen, a conditional latent-diffusion framework for generating paired H\&E counterfactuals with explicit controls over cellular spatial organization, nuclear morphology, and stain appearance. Cellular maps condition spatial structure through a spatial encoder, while a morphology/appearance vector modulates denoising through blockwise feature-wise linear modulation (FiLM). On held-out H\&E tiles, CPathOGen generates visually plausible tissue with improved distributional agreement after spatially guided selection, as reflected by lower Fréchet Inception Distance (FID) and Kernel Inception Distance (KID); generated cells track requested abundance and position, and measured morphology and color vary monotonically with their controls. We use these verified interventions to probe pathology encoders with endpoint heads, task-specific classifiers, and survival models. Responses are quantified using total variation distance and prediction-flip rate. We further introduce the Biology-Nuisance Sensitivity Ratio, a metric that contrasts model sensitivity to biologically motivated morphology and spatial factors with sensitivity to non-biological, stain-related nuisance variation. CPathOGen provides a practical, fidelity-audited framework for evaluating robustness and controlled feature sensitivity in computational pathology model. \href{https://github.com/a12dongithub/PathOGen}{GitHub} and \href{https://huggingface.co/a12donhf/CPathOGen}{Hugging~Face}.
\end{abstract}

\section{Introduction}
\label{sec:intro}

Histopathology provides a direct view of tissue morphology and is increasingly used to predict clinically meaningful outcomes~\cite{He2012, Shamai2026}. Computational pathology now includes models trained directly from whole-slide images (WSIs), often using only a slide-level diagnosis. Such models have been applied to cancer detection and subtyping, weakly supervised slide classification, and prediction of molecular biomarkers from H\&E tissue~\cite{campanella2019clinical,lu2021clam,kather2019msi,shmatko2022ai}. Large pretrained pathology encoders have widened the set of tasks that can be studied with limited labels. CTransPath uses contrastive pretraining; UNI and Virchow learn visual representations from large slide collections; CONCH pairs pathology images with text~\cite{wang2022ctranspath,chen2024uni,lu2024conch,vorontsov2024virchow}. Strong benchmark performance, however, does not establish how reliably these representations generalize across image sets or feature variations. The evidence used by these models is still difficult to interpret~\cite{Asif2023, https://doi.org/10.48550/arxiv.2510.23807}.

\begin{figure}[!t]
    \centering
    \includegraphics[width=\columnwidth]{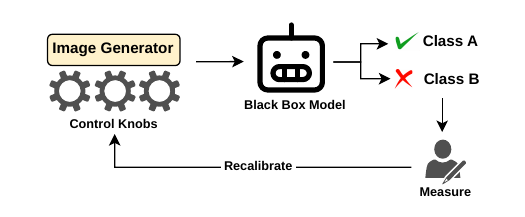}
    \vspace{-4mm}
    \caption{Underlying principle of post-hoc model interpretation using controllable counterfactual generation and measured prediction changes.}
    \label{fig:probing_cartoon}
    \vspace{-4mm}
\end{figure}

The appearance of an H\&E image reflects both the underlying tissue and how the slide was prepared and digitized. Features such as nuclear shape, cell density, and tumor--immune organization can also be influenced by staining intensity or to the hospital or laboratory where a slide was prepared. Howard \etal{} showed that standard normalization and augmentation do not fully remove clues about where a slide came from~\cite{howard2021sitespecific}. A model can therefore perform well on a held-out set partly by using clues from how the slides were prepared or scanned. This is one reason post-hoc explanations in medical AI require careful validation~\cite{ghassemi2021falsehope,klauschen2024xai}.

Saliency and attention maps provide spatial localization, but it is often unclear what features within a highlighted region influence the model's prediction. A tumor region may be influential because of nuclear size, immune density, spatial mixing, stain, or a scanner artifact. Adebayo \etal{} also showed that some saliency maps remain visually similar after randomizing model parameters or training labels~\cite{adebayo2018sanity}. In a pathology image, knowing where the model focuses does not reveal which tissue feature influenced its response.

Visual counterfactuals give a more direct test by modifying the input and measuring the resulting prediction. Goyal \etal{} retrieve discriminative changes relative to a target class~\cite{goyal2019counterfactual}. DVCE, DiME, and DiG-IN use diffusion priors to produce realistic classifier-guided edits~\cite{augustin2022dvce,jeanneret2022dime,augustin2024digin}. Pathology studies have used changes between classes to reveal tissue patterns linked to molecular labels~\cite{dolezal2023synthetic}. MoPaDi couples a diffusion autoencoder to a task-specific classifier or multiple-instance model to generate morphological counterfactuals optimized toward a target prediction~\cite{zigutyte2026mopadi}. Several correlated attributes may change during this optimization, making it difficult to attribute the model response to one requested factor.

\method{} begins with a requested intervention rather than a target prediction (Fig.~\ref{fig:probing_cartoon}). It treats cellular organization, nuclear morphology, and stain appearance as control knobs for conditional H\&E generation. Moving one knob produces matched counterfactuals while preserving the remaining modeled factors. After verifying the requested change, the images are passed through a black-box pathology model, and its prediction changes are recorded to characterize post-hoc feature sensitivity.

We evaluate image realism and verify that generated images reflect the requested interventions before downstream analysis. Weak cellular labels are extracted using \cellvit{}, while \hovernet{} and \stardist{} provide independent measurements~\cite{horst2026cellvitpp,graham2019hovernet,schmidt2018stardist}. The diffusion model was trained on 1.4 million $512\!\times\!512$ H\&E tiles sampled from 1,114 WSIs, one per patient. The verified counterfactuals are used to probe pathology models for PAM50 breast cancer subtype classification~\cite{Parker2009} and overall survival prediction.

The paper makes two technical contributions:
\begin{itemize}
    \item a trained diffusion model for controllable H\&E synthesis that reflects requested spatial, morphological, and appearance interventions; and
    \item an analysis of pathology foundation models and task-specific predictors using the verified counterfactuals to measure their responses to controlled feature changes.
\end{itemize}

\section{Related Work}
\label{sec:related}

\subsection{Explainability and counterfactual model probing}

Many post-hoc image explanations assign importance to pixels or regions. Their reliability depends on whether the explanation actually changes with the learned model, a point made explicit by saliency-map sanity checks~\cite{adebayo2018sanity} and by work on explainability in precision pathology~\cite{klauschen2024xai}. Counterfactual explanations take a different form: they search for a nearby input with a different prediction. Goyal \etal{} use discriminative visual comparisons~\cite{goyal2019counterfactual}. DVCE and DiME place classifier-guided edits under diffusion image priors~\cite{augustin2022dvce,jeanneret2022dime}, and DiG-IN extends diffusion guidance to comparisons between networks and neuron-level visualization~\cite{augustin2024digin}.

Interventions have also been used to examine the generative model itself. Tinaz \etal{} fit sparse autoencoders to diffusion activations, identify concepts at different denoising stages, and manipulate image composition and style through those concepts~\cite{tinaz2025concepts}. This line of work is useful for understanding where a generative representation forms. Our experiments instead use a conditional generator as an input instrument for testing external pathology models.

Histopathology counterfactual methods include synthetic class transitions linked to molecular labels~\cite{dolezal2023synthetic} and MoPaDi, which couples a diffusion autoencoder~\cite{preechakul2022diffae} to classifiers and multiple-instance models to edit morphology toward a target prediction~\cite{zigutyte2026mopadi}. These approaches motivate biologically plausible edits. Our evaluation additionally requires the requested intervention to be recovered from the generated image before that image is used for model probing.


\vspace{-6pt}
\subsection{Controllable histopathology generation}



PathologyGAN learned realistic tissue structure and interpretable latent representations using a generative adversarial model~\cite{quiros2021pathologygan}. PathLDM and PixCell operate in latent diffusion space: PathLDM introduced text-conditioned synthesis of histopathology patches, while PixCell scaled embedding-conditioned generation across tissue types and resolutions, with additional cell-mask guidance~\cite{yellapragada2024pathldm,https://doi.org/10.48550/arxiv.2506.05127}. Harb \etal{} extended diffusion synthesis to gigapixel WSIs~\cite{harb2024gigapixel}, and Spatial Diffusion modeled explicit cellular layouts~\cite{li2024spatialdiffusion}. TopoCellGen instead generates multiclass cell layouts under cell-count and topological constraints, using the resulting synthetic image-layout pairs to improve downstream cell detection and classification~\cite{Xu2025}. Collectively, these methods advance realism, scale, conditioning, or synthetic-data augmentation, but are not designed to provide separable controls over cellular organization, measured nuclear morphology, and appearance for matched single-factor interventions.

CPathOGen addresses this gap by combining a spatial encoder for cellular maps with FiLM-based conditioning of morphology and appearance~\cite{perez2018film}. This separation allows one control family to be varied while preserving the remaining modeled conditions, enabling direct measurement of whether each requested tissue change is faithfully expressed.




\begin{figure*}[!t]
    \centering
    \includegraphics[width=\textwidth,height=0.30\textheight,keepaspectratio]{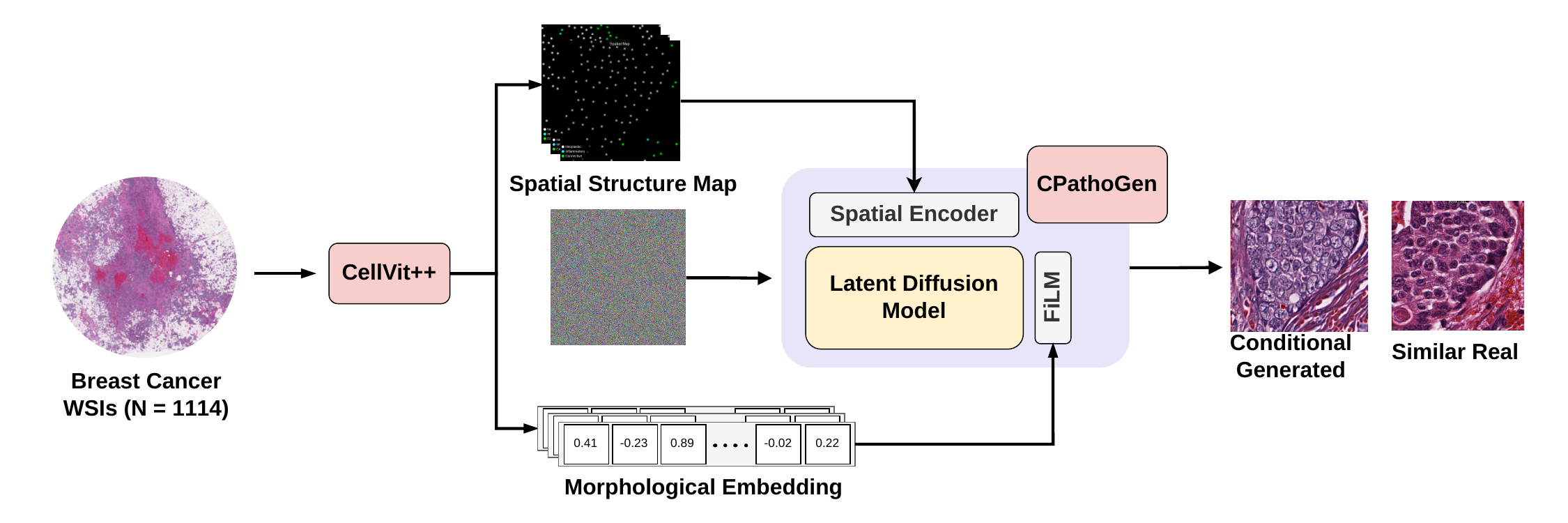}
    \vspace{-4mm}
    \caption{Overview of \method{} condition construction and synthesis. \cellvit{}-derived weak labels provide a cellular spatial map and a morphology/appearance embedding. The spatial encoder conditions the latent diffusion model on cellular organization, while FiLM modulates denoising with global morphology and stain attributes. These complementary controls enable realistic H\&E synthesis with separately specified spatial and morphological properties.}
    \label{fig:overview}
\end{figure*}

\section{Method}
\label{sec:method}

\subsection{Framework overview}
\label{sec:framework-overview}

CPathOGen is an interventional image-synthesis framework for examining how computational pathology models respond to controlled changes in H\&E input tiles. Figure~\ref{fig:probing_cartoon} presents the overall use case: a control is modified, CPathOGen generates a matched H\&E counterfactual, and the response of a frozen downstream model is recorded for model interpretation.

Figure~\ref{fig:overview} summarizes the construction of the conditional generator. During training, CellViT++ extracts weak cellular labels from real H\&E tiles. These labels are converted into a cellular spatial map and a global morphology/appearance vector, which condition a Stable Diffusion 2.1 latent-diffusion model through latent spatial conditioning and blockwise FiLM. At inference, the same condition format can be extracted from a real tile, manually modified, or constructed by the user.

\subsection{Weakly supervised condition construction}
\label{sec:weak-conditions}

As illustrated on the left of Figure~\ref{fig:overview}, to construct conditions for generator training at scale, we use pretrained nucleus segmentation and cell-classification models. In our implementation, CellViT++ provides nucleus contours, centroids, and predicted cell types. These predictions are treated as weak labels rather than manual ground truth.

Let an analyzer $\mathcal{A}$ applied to an H\&E tile $x$ produce
\begin{equation}
\mathcal{A}(x)
=
\left\{
    (\mathbf{q}_n,y_n,\omega_n)
\right\}_{n=1}^{N},
\label{eq:nucleus-set}
\end{equation}
where $N$ is the number of detected nuclei, $\mathbf{q}_n$ is the centroid of nucleus $n$, $y_n$ is its predicted cellular compartment, and $\omega_n$ is its instance contour.

The spatial condition is represented by
$S\in[0,1]^{K\times H\times W}$, where $K$ is the number of cellular compartments and $H$ and $W$ are the tile height and width. Each channel is constructed by rasterizing the centroids belonging to one cellular compartment:
\begin{equation}
S_k(\mathbf{u})
=
\operatorname{Norm}
\left[
    \sum_{n:y_n=k}
    \kappa(\mathbf{u}-\mathbf{q}_n)
\right],
\label{eq:spatial-map}
\end{equation}
where $\mathbf{u}$ denotes a pixel location, $\kappa$ is a localized spatial kernel, and $\operatorname{Norm}$ scales the resulting channel to a common range. We use $K=5$ channels corresponding to tumor, inflammatory, epithelial, stromal, and necrotic compartments.

A global morphology and appearance condition
$m\in\mathbb{R}^{d}$ is computed from the detected nucleus contours and corresponding image regions:
\begin{equation}
m
=
\operatorname{Std}
\left[
    \Phi\bigl(x,\mathcal{A}(x)\bigr)
\right],
\label{eq:morph-vector}
\end{equation}
where $\Phi$ aggregates the mean and dispersion of nuclear size, perimeter, eccentricity, solidity, nuclear gradient, and RGB appearance statistics. The operator $\operatorname{Std}$ applies the feature-wise standardization used during model training. Our implementation uses $d=16$ features.

The preprocessing parameters, feature ordering, and supported control ranges are stored with the model and reused during inference. This is particularly important for the appearance dimensions, for which incorrect standardization or feature ordering can produce globally inconsistent H\&E coloration.

\subsection{Conditional latent-diffusion model}
\label{sec:conditional-ldm}

CPathOGen as shown in Figure~\ref{fig:overview} is built upon a latent diffusion model initialized from pretrained Stable Diffusion 2.1 weights~\cite{rombach2022latent}. We first adapt the pretrained visual prior to H\&E tissue and subsequently introduce the structured spatial and morphology conditions.

Let $\mathcal{E}$ and $\mathcal{D}$ denote the encoder and decoder of the pretrained variational autoencoder. An H\&E tile $x$ is encoded as
\begin{equation}
z_0=\mathcal{E}(x),
\end{equation}
where $z_0$ is its lower-dimensional latent representation. At diffusion step $t$, Gaussian noise $\epsilon$ is added according to
\begin{equation}
z_t
=
\sqrt{\bar{\alpha}_t}\,z_0
+
\sqrt{1-\bar{\alpha}_t}\,\epsilon,
\qquad
\epsilon\sim\mathcal{N}(0,I),
\label{eq:forward-diffusion}
\end{equation}
where $\bar{\alpha}_t$ controls the amount of signal retained at step $t$.

The conditional denoising U-Net $\epsilon_\theta$ predicts the added noise from the noisy latent, diffusion timestep, spatial condition, and morphology vector. It is optimized using
\begin{equation}
\mathcal{L}_{\mathrm{diff}}
=
\mathbb{E}_{x,t,\epsilon}
\left[
    \left\|
        \epsilon-
        \epsilon_\theta(z_t,t;S,m)
    \right\|_2^2
\right].
\label{eq:diffusion-loss}
\end{equation}
The text condition is held fixed because synthesis is governed by the structured pathology controls rather than natural-language prompts.

\subsection{Latent spatial conditioning}
\label{sec:spatial-conditioning}

The spatial condition is introduced through a trainable convolutional encoder
$E_s$. Given the cellular map
$S\in[0,1]^{K\times H\times W}$, the encoder produces
\begin{equation}
c_s=E_s(S),
\qquad
c_s\in\mathbb{R}^{C_s\times h\times w},
\label{eq:spatial-encoder}
\end{equation}
where $C_s$ is the number of spatial feature channels and $h\times w$ is the spatial resolution of the diffusion latent. In our implementation,
$H=W=512$, $h=w=64$, and $C_s=4$.

Let
$z_t\in\mathbb{R}^{C_z\times h\times w}$ denote the noisy image latent at diffusion step $t$, with $C_z=4$. The encoded spatial features are concatenated with the noisy latent:
\begin{equation}
\widetilde{z}_t
=
\left[
    z_t;\lambda_s c_s
\right]
\in
\mathbb{R}^{(C_z+C_s)\times h\times w},
\label{eq:latent-concatenation}
\end{equation}
where $[\cdot;\cdot]$ denotes channel-wise concatenation and $\lambda_s$ controls the strength of spatial conditioning. The resulting tensor contains eight channels in our implementation.

The first U-Net convolution is expanded from four to eight input channels. The weights associated with the original latent channels retain their pretrained initialization, while the additional spatial-input weights are initialized to zero. The model can therefore begin from the pretrained H\&E image prior and progressively learn how cell type, abundance, and location should affect synthesis.

\subsection{Morphology and Stain modulation}
\label{sec:film-conditioning}

The global condition $m$ is injected throughout the denoising U-Net using feature-wise linear modulation. Let
\[
h_\ell
\in
\mathbb{R}^{B\times C_\ell\times h_\ell\times w_\ell}
\]
denote the output activation of residual block $\ell$, where $B$ is the batch size, $C_\ell$ is the number of channels, and $h_\ell\times w_\ell$ is the spatial resolution at that block.

Each residual block has a learned two-layer mapping
\[
f_\ell:\mathbb{R}^{d}\rightarrow\mathbb{R}^{2C_\ell},
\]
which transforms the morphology vector into channel-wise scale and shift parameters:
\begin{equation}
(\gamma_\ell,\beta_\ell)=f_\ell(m).
\label{eq:film-parameters}
\end{equation}
Here,
$\gamma_\ell,\beta_\ell\in\mathbb{R}^{B\times C_\ell}$ are broadcast over the spatial dimensions of $h_\ell$. The modulated activation is
\begin{equation}
h'_\ell
=
(1+\gamma_\ell)\odot h_\ell+\beta_\ell,
\label{eq:film}
\end{equation}
where $\odot$ denotes channel-wise multiplication. The identity offset in $1+\gamma_\ell$ preserves the unmodulated block when the predicted scale is zero.

Applying FiLM across the U-Net residual blocks allows nuclear geometry, texture, and appearance to influence multiple levels of the denoising hierarchy. The scale and shift parameters are bounded to prevent extreme conditioning values from overwhelming the learned image representation. The spatial map and morphology vector consequently provide complementary controls: $S$ determines where cellular structures should occur, while $m$ controls their aggregate geometry and appearance.

\subsection{Prediction-independent fidelity guidance}
\label{sec:fidelity-guidance}

Different diffusion seeds can vary in how faithfully they express the requested cellular arrangement. For each condition, we therefore generate a fixed set of candidate images and rank them according to agreement between the requested spatial map and the cellular organization recovered from each generated image.

Let $\mathcal{R}$ denote the candidate seed set and
\[
x_r=G_\theta(S,m,\epsilon_r)
\]
the image generated using seed $r\in\mathcal{R}$. A cellular analyzer is applied to $x_r$, and a fidelity function $Q(x_r,S)$ measures agreement between requested and recovered cell types and locations. The selected seed is
\begin{equation}
r^\star
=
\arg\max_{r\in\mathcal{R}}
Q(x_r,S).
\label{eq:candidate-selection}
\end{equation}
CellViT++ supplies the primary agreement signal in our implementation. The fidelity function and candidate set are fixed before downstream evaluation, and selection does not observe the predictions, representations, or gradients of the model being probed. The selected image is then used for the corresponding counterfactual analysis.

\subsection{Paired interventions and model probing}
\label{sec:paired-interventions}

At inference, users may edit an extracted condition or directly construct a custom spatial map and morphology vector. Let
$c=(S,m)$ denote a baseline condition and
$\mathcal{I}_{j,\delta}$ an intervention of magnitude $\delta$ applied to control family $j$. A matched image pair is generated as
\begin{equation}
x^{(0)}
=
G_\theta(S,m,\epsilon),
\qquad
x^{(j,\delta)}
=
G_\theta\left(
    \mathcal{I}_{j,\delta}(S,m),
    \epsilon
\right),
\label{eq:counterfactual-pair}
\end{equation}
where $G_\theta$ includes latent denoising and image decoding. Both images reuse the same initial noise $\epsilon$, and all non-target conditioning variables remain fixed.

For example, morphology interventions vary one nuclear property while preserving spatial organization and appearance. Spatial interventions modify cell composition, density, distance, or mixing while preserving the global morphology vector. Appearance interventions modify the selected color controls while preserving spatial and geometric conditions. Reusing the same noise reduces unrelated stochastic variation, although the requested intervention is subsequently verified from the generated image.

Given a frozen downstream model $F$, its responses are
\begin{equation}
o^{(0)}=F(x^{(0)}),
\qquad
o^{(j,\delta)}=F(x^{(j,\delta)}),
\label{eq:probe-output}
\end{equation}
where $o$ may represent biological endpoint probabilities, a survival-risk estimate, or an image embedding. For every intervention, CPathOGen records the source tile, conditioning variables, intervention magnitude, diffusion seed, generated image, and downstream output. These matched intervention records support model interpretation by revealing which controlled biological or appearance changes alter the model's predictions or representations.
\section{Experiments}
\label{sec:experiments}

\subsection{Experimental setup}
\label{sec:experimental-setup}

We use 1,114 whole-slide images from The Cancer Genome Atlas (TCGA) Breast Cancer Cohort~\cite{tcga2013pancancer}. Slides are processed at
$20\times$ magnification and divided into $512\times512$ H\&E tiles.
CellViT++ provides weak nucleus contours, centroids, and cell-type labels.
Tiles containing more than two detected tumor cells are retained, yielding
approximately 1.4 million tiles with aligned spatial maps and
morphology/appearance vectors. Dataset partitions are separated at the patient
level. CPathOGen is trained on eight NVIDIA Tesla V100 GPUs and evaluated using
a fixed checkpoint.

We use $N_R=10{,}000$ samples per distribution for realism and $N=1{,}000$
baseline conditions for the remaining experiments. Morphology and appearance
interventions use
\begin{equation}
\mathcal{D}=\{-2,-1,0,+1,+2\},
\label{eq:dose-set}
\end{equation}
where each value denotes the requested displacement from the baseline in
training-set standard deviations. Thus, $-2$ and $+2$ represent two standard
deviations below and above baseline, while $0$ leaves the condition unchanged.
Spatial experiments instead use ordered changes expressed directly in the
cellular map.

ResNet-50~\cite{https://doi.org/10.48550/arxiv.1512.03385}, CTransPath, and UNI2-h are used as frozen feature extractors, and
their tile embeddings are $\ell_2$-normalized. For PAM50, patient- and
class-balanced multinomial logistic regression heads are trained at the tile
level under five patient-disjoint folds. Patient-level performance is obtained
by averaging tile probabilities within each patient and is reported using macro
one-vs-rest AUC. Counterfactual responses are measured directly between paired
baseline and intervention tiles.


Overall survival is evaluated using a fixed 16-tile bag for each patient. This common tile budget enables a consistent comparison between frozen embedding models and PathLUPI, a fully pretrained survival model designed natively for multi-tile input. For ResNet-50, CTransPath, and UNI2-h, tile embeddings are mean-pooled and $\ell_2$-normalized before fitting a censor-aware Cox head. Each counterfactual replaces one tile in an otherwise unchanged bag. PathLUPI~\cite{https://doi.org/10.48550/arxiv.2506.19681} receives the same 16 tiles but retains its released CONCH-based aggregation and survival formulation. Performance is measured using Harrell's C-index, while counterfactual responses are calculated from predicted five-year survival probabilities. Generated images are never used for downstream-model training.

\subsection{Distributional realism}
\label{sec:distributional-realism}

FID~\cite{https://doi.org/10.48550/arxiv.1706.08500} and KID~\cite{https://doi.org/10.48550/arxiv.1801.01401} compare $N_R$ generated images with an independent reference set of
$N_R$ real tiles. Generation conditions are obtained from a separate,
non-overlapping set of real tiles, ensuring that the conditioning and reference
distributions contain no shared images. The unfiltered setting uses one
generation per condition, whereas the selected setting applies the
eight-candidate spatial selection procedure from Sec.~\ref{sec:method}.

\begin{table}[t]
\centering
\caption{Distributional realism using $N_R=10{,}000$ generated images and an
independent real reference set of equal size.}
\label{tab:realism}
\setlength{\tabcolsep}{3.5pt}
\begin{tabular}{lcc}
\toprule
Evaluation set & FID $\downarrow$ & KID $\downarrow$ \\
\midrule
No seed filtering
& 42.923
& 0.02466 \\
CellViT++ selected
& \textbf{36.717}
& \textbf{0.01964} \\
Real vs.\ real
& 16.369
& 0.00647 \\
\bottomrule
\end{tabular}
\end{table}

\begin{figure}[!t]

\centering

\scriptsize

\setlength{\tabcolsep}{1pt}

\begin{tabular}{@{}ccc@{}}

\textbf{Cell map} & \textbf{Real H\&E} & \textbf{Generated H\&E} \\[-1pt]

\includegraphics[width=0.28\columnwidth]
{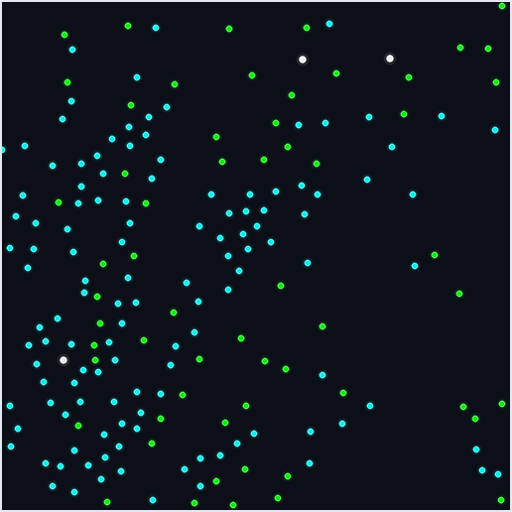}
&
\includegraphics[width=0.28\columnwidth]
{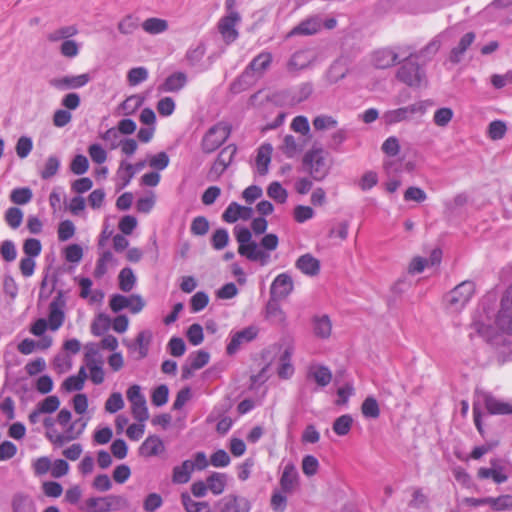}
&
\includegraphics[width=0.28\columnwidth]
{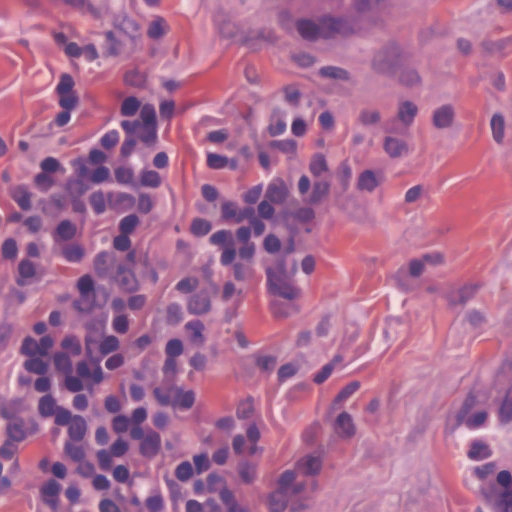}
\\[1pt]

\includegraphics[width=0.28\columnwidth]
{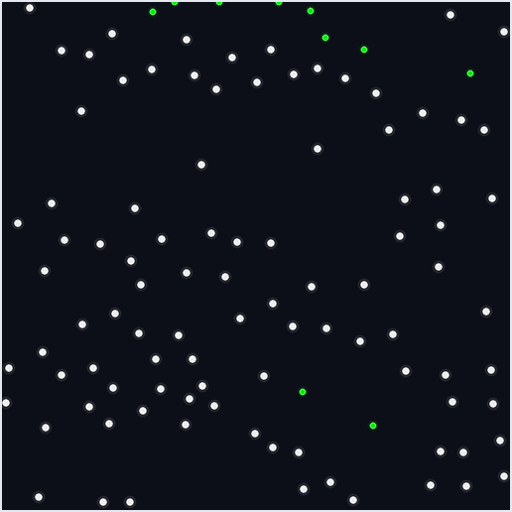}
&
\includegraphics[width=0.28\columnwidth]
{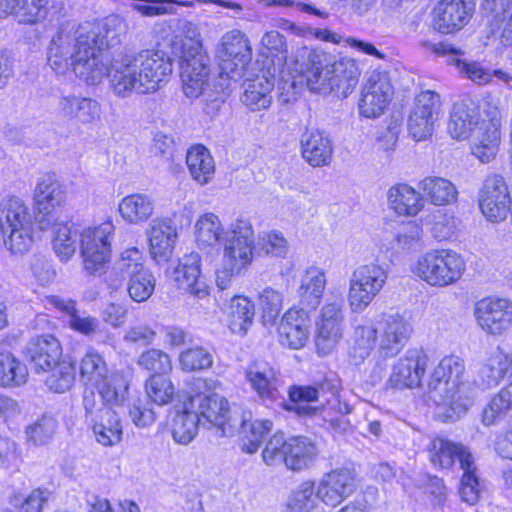}
&
\includegraphics[width=0.28\columnwidth]
{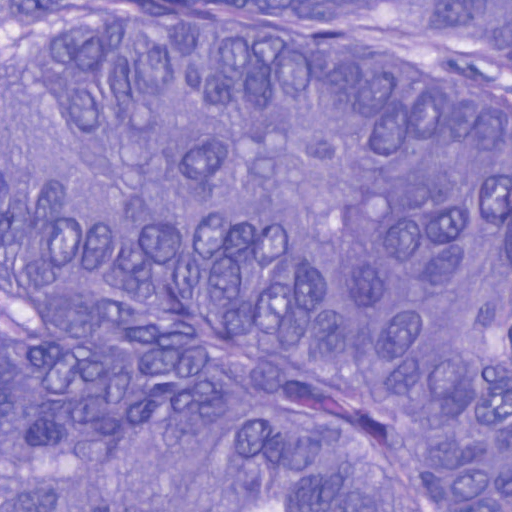}
\\[1pt]

\includegraphics[width=0.28\columnwidth]
{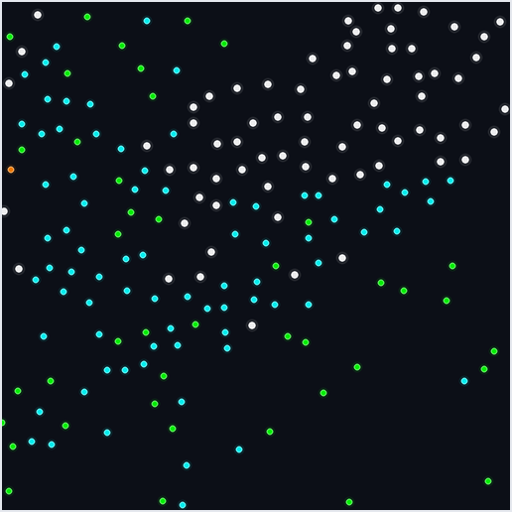}
&
\includegraphics[width=0.28\columnwidth]
{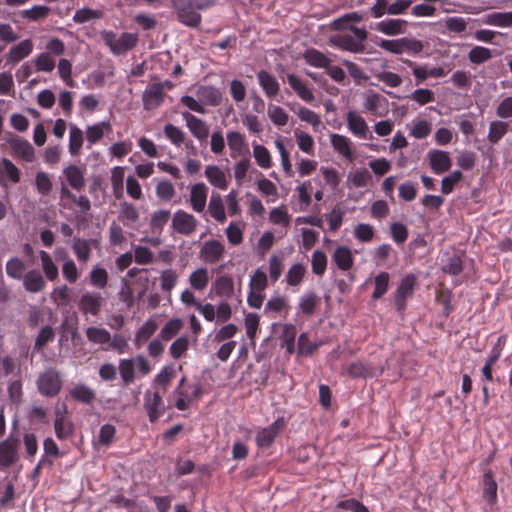}
&
\includegraphics[width=0.28\columnwidth]
{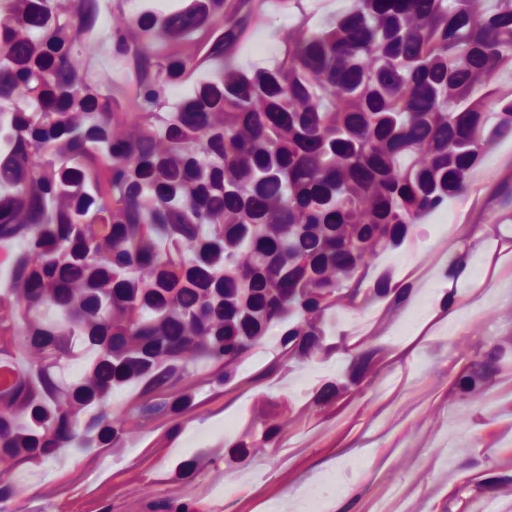}
\\

\end{tabular}

\par\vspace{2pt}

\begingroup

\definecolor{mapcyan}{RGB}{0,255,255}
\definecolor{mapgreen}{RGB}{0,255,0}
\definecolor{mapyellow}{RGB}{255,255,0}
\definecolor{maporange}{RGB}{255,128,0}

\setlength{\fboxsep}{0pt}

\newcommand{\mapkey}[2]{%
  \raisebox{-0.25ex}{%
    \fcolorbox{black}{#1}{%
      \rule{0pt}{1.05ex}\rule{1.05ex}{0pt}%
    }%
  }\,#2%
}

\mapkey{white}{Tumor}\quad
\mapkey{mapcyan}{Immune}\quad
\mapkey{mapgreen}{Stroma}\quad
\mapkey{mapyellow}{Dead}\quad
\mapkey{maporange}{Non-neoplastic epithelium}

\endgroup

\caption{Representative spatial maps with their corresponding real and
generated H\&E tiles. Generated images follow the requested organization
without reconstructing the exact real reference.}

\label{fig:spatial_grid}

\end{figure}

Spatially guided selection improves both metrics, indicating that candidates
which better express the requested cellular structure also more closely follow
the held-out H\&E distribution. The real-versus-real row contextualizes
residual finite-sample variation. Figure~\ref{fig:spatial_grid} complements
these distributional metrics by showing real and generated tiles side by side
under the same spatial and morphology conditions. The generated tile is not
expected to reproduce the real tile pixelwise, but provides a qualitative
assessment of realism under matched conditioning.

We avoid comparison with values from unrelated pathology datasets because FID
and KID depend strongly on magnification, sampling, preprocessing, and
reference composition.

\subsection{Spatial fidelity}
\label{sec:spatial-fidelity}

Spatial fidelity measures whether the cellular abundance and organization
specified by an input map are recovered from the generated tile. Total-count
fidelity is the Spearman correlation between requested and recovered cellular
burden over $N$ conditions. Per-type fidelity is calculated for tumor, immune,
stromal, necrotic, and non-neoplastic epithelial cells and macro-averaged.

\begin{table}[!t]
\centering
\caption{Spatial fidelity over $N=1{,}000$ baselines. Spearman $\rho$
measures requested-versus-recovered cell counts; centroid F1 uses one-to-one
matching at 25/50 px. Bold denotes the best result.}
\label{tab:spatial}
\scriptsize
\renewcommand{\arraystretch}{1.12}
\setlength{\tabcolsep}{2pt}

\begin{tabularx}{\columnwidth}{
    @{}l
    *{4}{>{\centering\arraybackslash}X}
    @{}
}
\toprule
Method
& \shortstack{Total count\\$\rho\uparrow$}
& \shortstack{Per-type\\count $\rho\uparrow$}
& \shortstack{Centroid F1\\@ 25 px $\uparrow$}
& \shortstack{Centroid F1\\@ 50 px $\uparrow$} \\
\midrule
CellViT++
& \textbf{0.952}
& \textbf{0.612}
& 0.726
& 0.773 \\
HoVer-Net
& 0.908
& 0.525
& 0.625
& 0.668 \\
StarDist
& 0.928
& N/A
& \textbf{0.798}
& \textbf{0.831} \\
Random real tile
& $-0.031$
& $-0.020$
& 0.134
& 0.255 \\
\bottomrule
\end{tabularx}
\end{table}

Localization uses one-to-one Hungarian centroid matching at 25 and 50 pixels
at the $20\times$ (~0.5 µm per pixel) evaluation resolution. CellViT++ and HoVer-Net require
agreement in position and type, while StarDist evaluates position only. A
random-real-tile control pairs each map with a tile from another patient,
providing realistic tissue without correspondence to the requested map.

All three analyzers recover total cellular burden consistently. Per-type
agreement is lower because it additionally depends on cell classification and
agreement between analyzer taxonomies. Sparsity of rare classes, particularly
necrotic cells, further reduces the stability of type-specific correlations.
The random-tile control remains substantially weaker, showing that agreement
is driven by the requested map rather than the presence of plausible nuclei
alone.

StarDist achieves the highest position-only F1 but is not penalized for type
disagreement. Agreement across CellViT++, HoVer-Net, and StarDist nevertheless
shows that spatial fidelity is not specific to the selection model.
Figure~\ref{fig:spatial_grid} provides corresponding qualitative examples.

\begin{table*}[!t]
\centering
\caption{Morphology and appearance fidelity over $N=1{,}000$ baselines.
Across-image rows report Spearman $\rho$ across baselines; controlled rows
report the median within-source $\rho$ over $\mathcal{D}$. Macro averages
attributes; bold denotes the best result per block.}
\label{tab:morphology}
\footnotesize
\renewcommand{\arraystretch}{0.92}
\setlength{\tabcolsep}{4pt}

\begin{tabular*}{\textwidth}{
  @{\extracolsep{\fill}}l*{8}{c}@{}
}
\toprule
Method
& Size
& Eccentricity
& Solidity
& Gradient
& Red
& Green
& Blue
& Macro \\
& $\rho\uparrow$
& $\rho\uparrow$
& $\rho\uparrow$
& $\rho\uparrow$
& $\rho\uparrow$
& $\rho\uparrow$
& $\rho\uparrow$
& $\rho\uparrow$ \\
\midrule
Across images CellViT++
& \textbf{0.941}
& \textbf{0.820}
& \textbf{0.649}
& 0.878
& \textbf{0.951}
& \textbf{0.961}
& \textbf{0.928}
& \textbf{0.876} \\
Across images HoVer-Net
& 0.896
& 0.798
& 0.499
& 0.886
& 0.942
& 0.958
& 0.925
& 0.843 \\
Across images StarDist
& 0.935
& 0.760
& 0.570
& \textbf{0.892}
& 0.950
& 0.954
& \textbf{0.928}
& 0.855 \\
\midrule
Controlled CellViT++
& \textbf{1.000}
& \textbf{1.000}
& \textbf{0.900}
& \textbf{1.000}
& \textbf{1.000}
& \textbf{1.000}
& \textbf{1.000}
& \textbf{0.986} \\
Controlled HoVer-Net
& \textbf{1.000}
& \textbf{1.000}
& \textbf{0.900}
& \textbf{1.000}
& \textbf{1.000}
& \textbf{1.000}
& \textbf{1.000}
& \textbf{0.986} \\
Controlled StarDist
& \textbf{1.000}
& \textbf{1.000}
& 0.500
& \textbf{1.000}
& \textbf{1.000}
& \textbf{1.000}
& \textbf{1.000}
& 0.929 \\
\bottomrule
\end{tabular*}
\end{table*}

\begin{figure*}[!t]
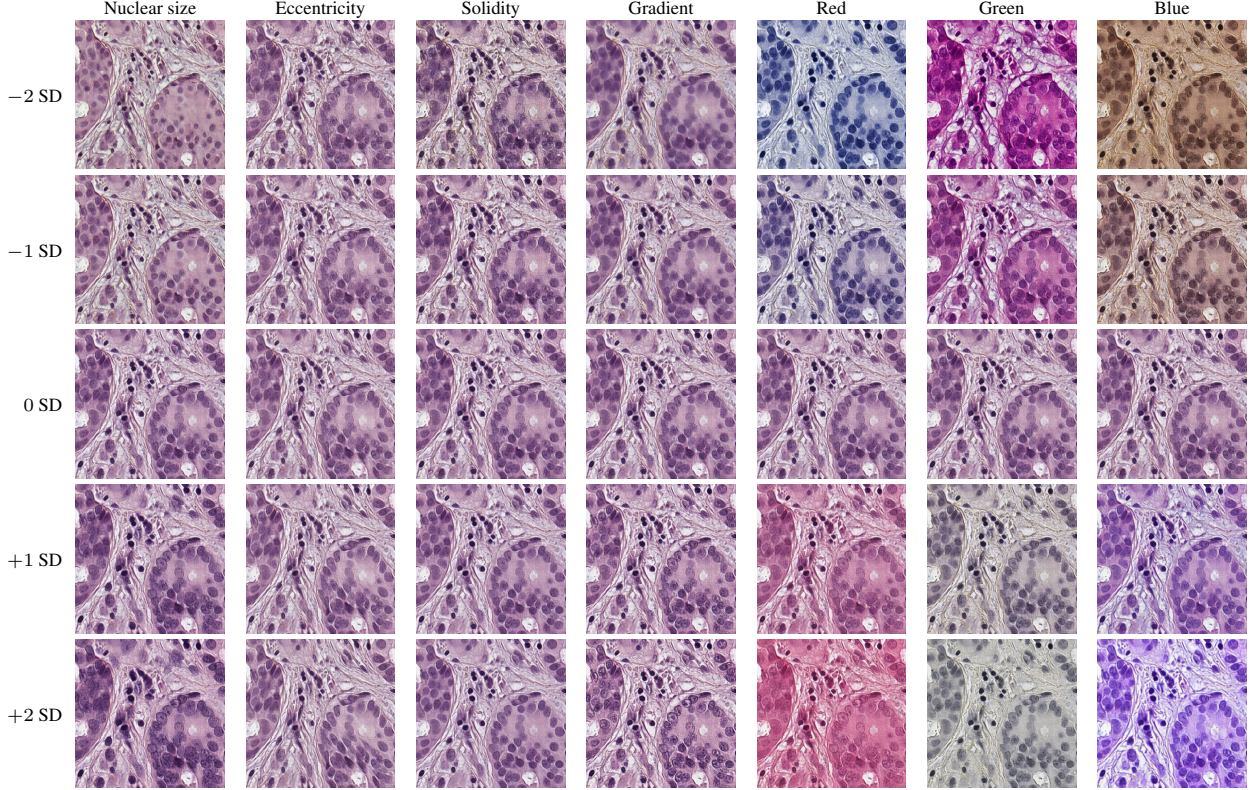

\centering
\scriptsize
\setlength{\tabcolsep}{1pt}
\renewcommand{\arraystretch}{0.98}

\newcommand{\morphtile}[2]{%
  \includegraphics[width=0.9\linewidth]
  {01_morphology_controls/#1/#2.png}%
}

\begin{tabular}{@{}
  >{\raggedleft\arraybackslash}m{0.075\textwidth}
  *{7}{>{\centering\arraybackslash}m{0.125\textwidth}}
  @{}}

& Nuclear size
& Eccentricity
& Solidity
& Gradient
& Red
& Green
& Blue \\

$-2$ SD
& \morphtile{nuclear_size}{minus_2sd}
& \morphtile{eccentricity}{minus_2sd}
& \morphtile{solidity}{minus_2sd}
& \morphtile{gradient}{minus_2sd}
& \morphtile{red}{minus_2sd}
& \morphtile{green}{minus_2sd}
& \morphtile{blue}{minus_2sd}
\\[1pt]

$-1$ SD
& \morphtile{nuclear_size}{minus_1sd}
& \morphtile{eccentricity}{minus_1sd}
& \morphtile{solidity}{minus_1sd}
& \morphtile{gradient}{minus_1sd}
& \morphtile{red}{minus_1sd}
& \morphtile{green}{minus_1sd}
& \morphtile{blue}{minus_1sd}
\\[1pt]

$0$ SD
& \morphtile{nuclear_size}{zero_sd}
& \morphtile{eccentricity}{zero_sd}
& \morphtile{solidity}{zero_sd}
& \morphtile{gradient}{zero_sd}
& \morphtile{red}{zero_sd}
& \morphtile{green}{zero_sd}
& \morphtile{blue}{zero_sd}
\\[1pt]

$+1$ SD
& \morphtile{nuclear_size}{plus_1sd}
& \morphtile{eccentricity}{plus_1sd}
& \morphtile{solidity}{plus_1sd}
& \morphtile{gradient}{plus_1sd}
& \morphtile{red}{plus_1sd}
& \morphtile{green}{plus_1sd}
& \morphtile{blue}{plus_1sd}
\\[1pt]

$+2$ SD
& \morphtile{nuclear_size}{plus_2sd}
& \morphtile{eccentricity}{plus_2sd}
& \morphtile{solidity}{plus_2sd}
& \morphtile{gradient}{plus_2sd}
& \morphtile{red}{plus_2sd}
& \morphtile{green}{plus_2sd}
& \morphtile{blue}{plus_2sd}
\\
\end{tabular}

\caption{Controlled morphology and appearance interventions. Each column
varies one feature over $\mathcal{D}$ while preserving the spatial map,
diffusion seed, and all other morphology/appearance entries.}
\label{fig:morphology_grid}
\end{figure*}

\subsection{Morphology and appearance fidelity}
\label{sec:morphology-fidelity}

Across-image fidelity correlates requested and recovered features over the $N$
independent baseline conditions. Within-image fidelity varies one feature over
$\mathcal{D}$ while preserving the spatial map, diffusion seed, and all other
morphology/appearance entries. A separate Spearman correlation is calculated
across doses for each source and feature.

Nuclear size, eccentricity, and solidity are measured from predicted contours,
while gradient and RGB values are measured within detected nuclear regions.
Perimeter remains part of the conditioning vector but is omitted because of
its strong relationship with area.

Across-image fidelity uses one cohort-level correlation because each image
contributes one independent pair. Within-image correlations are calculated per
source and summarized by their median, preventing baseline morphology from
confounding the controlled response. Macro $\rho$ averages features equally.

Agreement is strongest for size, gradient, and color and remains consistent
across independent analyzers. The strong within-image results demonstrate that
recovered features generally follow the requested ordering when a single
control is varied. Figure~\ref{fig:morphology_grid}
shows the corresponding visual progression.

\begin{table*}[t]
\centering
\caption{
Predictive performance and counterfactual responses.
Intervention entries report mean TVD / prediction-flip rate.
Within each model, bold and underlined intervention values denote the highest
and lowest responses, respectively, evaluated separately for TVD and flip rate.
BNR compares mean sensitivity to biological interventions with mean sensitivity
to the stain and rotation nuisance controls; brackets denote patient-clustered
95\% bootstrap confidence intervals.
}
\label{tab:xai_all_tasks}

\small
\renewcommand{\arraystretch}{1.25}
\setlength{\tabcolsep}{3.5pt}

\begin{adjustbox}{max width=\textwidth}
\begin{tabular}{lcccccccc}
\toprule
\textbf{Model}
& \textbf{Performance}
& \shortstack{\textbf{Stain}\\\textbf{brightness}}
& \shortstack{\textbf{Image}\\\textbf{rotation}}
& \shortstack{\textbf{Nuclear}\\\textbf{enlargement}}
& \shortstack{\textbf{Shape}\\\textbf{irregularity}}
& \shortstack{\textbf{Immune}\\\textbf{burden}}
& \shortstack{\textbf{Tumor--immune}\\\textbf{mixing}}
& \shortstack{\textbf{BNR}\\\textbf{[95\% CI]}} \\
\midrule

\multicolumn{1}{l}{\textbf{PAM50 classification}}
& \textit{AUC $\uparrow$}
& \multicolumn{7}{c}{} \\
\midrule

ResNet-50
& 0.7577
& 0.1687 / 0.3859
& 0.1203 / 0.2867
& 0.1647 / 0.3638
& \textbf{0.1759} / \textbf{0.4041}
& 0.1672 / 0.3474
& \underline{0.1126} / \underline{0.2482}
& \shortstack{1.0736\\{\scriptsize [1.0493, 1.0975]}} \\

CTransPath
& 0.8323
& \textbf{0.2181} / \textbf{0.3670}
& \underline{0.0846} / \underline{0.1483}
& 0.1883 / 0.3157
& 0.1852 / 0.3071
& 0.1993 / 0.3042
& 0.1194 / 0.1995
& \shortstack{1.1434\\{\scriptsize [1.1154, 1.1737]}} \\

UNI2-h
& 0.8687
& 0.2456 / 0.3511
& \underline{0.1089} / \underline{0.1540}
& \textbf{0.2596} / \textbf{0.3628}
& 0.2321 / 0.3444
& 0.2567 / 0.3624
& 0.1582 / 0.2432
& \shortstack{1.2787\\{\scriptsize [1.2466, 1.3121]}} \\

\midrule

\multicolumn{1}{l}{\textbf{Overall survival}}
& \textit{C-index $\uparrow$}
& \multicolumn{7}{c}{} \\
\midrule

ResNet-50
& 0.5575
& 0.0094 / 0.0033
& 0.0094 / 0.0038
& 0.0093 / 0.0023
& 0.0114 / \textbf{0.0046}
& \textbf{0.0120} / 0.0036
& \underline{0.0075} / \underline{0.0013}
& \shortstack{1.0732\\{\scriptsize [1.0253, 1.1216]}} \\

CTransPath
& 0.6549
& 0.0057 / \textbf{0.0053}
& 0.0042 / \underline{0.0024}
& 0.0059 / 0.0033
& 0.0057 / 0.0028
& \textbf{0.0064} / 0.0047
& \underline{0.0037} / 0.0029
& \shortstack{1.0899\\{\scriptsize [1.0416, 1.1465]}} \\

UNI2-h
& 0.6691
& 0.0052 / 0.0020
& 0.0036 / \underline{0.0007}
& 0.0053 / \textbf{0.0047}
& 0.0052 / 0.0046
& \textbf{0.0058} / 0.0042
& \underline{0.0034} / 0.0026
& \shortstack{1.1131\\{\scriptsize [1.0680, 1.1610]}} \\

PathLUPI + CONCH
& 0.5977
& 0.0084 / 0.0107
& \underline{0.0040} / \underline{0.0027}
& 0.0090 / 0.0094
& 0.0094 / 0.0080
& \textbf{0.0201} / \textbf{0.0130}
& 0.0052 / 0.0055
& \shortstack{1.7572\\{\scriptsize [1.6132, 1.9202]}} \\

\bottomrule
\end{tabular}
\end{adjustbox}
\end{table*}

\subsection{Counterfactual model probing}
\label{sec:counterfactual-probing}
We evaluate five CPathOGen interventions and one generator-independent
nuisance control. Stain brightness jointly modifies RGB appearance. Nuclear
enlargement changes nuclear area and perimeter, while shape irregularity
changes eccentricity and reduces solidity while preserving nuclear area.
Immune burden changes inflammatory-cell abundance, and tumor--immune mixing
relocates immune cells from mixed to segregated configurations without changing
tumor or immune counts. Image rotation applies $0^\circ$, $90^\circ$,
$180^\circ$, and $270^\circ$ transformations to the same generated reference
tile.

We group nuclear enlargement, shape irregularity, immune burden, and
tumor--immune mixing into the biological intervention set $\mathcal{B}$,
reflecting nuclear phenotype and tumor-microenvironment organization
\cite{zink2004nuclear, Williams2024}. Stain brightness and image rotation
form the nuisance set $\mathcal{U}$, representing appearance and orientation
changes that should not independently determine the biological endpoint
\cite{tellez2019stain}. These groups define the Biology-to-Nuisance Ratio and
do not imply biological causality. Supplementary material presents
representative images from the five CPathOGen interventions.

For model $f$, let $\mathbf{p}_{f}(x_i)$ and
$\mathbf{p}_{f}(x_{i,e,\delta})$ denote the output distributions for baseline
$i$ and experiment $e$ at level $\delta$. Total variation distance is
\begin{equation}
d_{i,e,\delta}
=
\frac{1}{2}
\left\|
\mathbf{p}_{f}(x_{i,e,\delta})
-
\mathbf{p}_{f}(x_i)
\right\|_1 .
\label{eq:tvd}
\end{equation}
For PAM50, $\mathbf{p}_f$ is the four-class subtype distribution. For survival,
it is the Bernoulli distribution formed from the model's survival probability.
Prediction-flip rate records a change in the PAM50 argmax class or crossing the
$0.5$ survival-probability threshold.

Let $D_f(e)$ be the mean TVD over the $N$ baselines and evaluated non-reference
levels $\mathcal{D}_e$:
\begin{equation}
D_f(e)=
\frac{1}{N|\mathcal{D}_e|}
\sum_{i=1}^{N}
\sum_{\delta\in\mathcal{D}_e}
d_{i,e,\delta}.
\label{eq:experiment-response}
\end{equation}
The Biology-to-Nuisance Ratio is
\begin{equation}
\mathrm{BNR}(f)=
\frac{
|\mathcal{B}|^{-1}\sum_{e\in\mathcal{B}}D_f(e)
}{
|\mathcal{U}|^{-1}\sum_{e\in\mathcal{U}}D_f(e)
}.
\label{eq:bnr}
\end{equation}
BNR above one indicates a larger average response to the evaluated biological
interventions than to the stain and rotation nuisance controls. It measures
relative sensitivity, not predictive performance or biological correctness.

UNI2-h achieves the strongest PAM50 discrimination, but it is not the least
sensitive model, demonstrating that predictive performance and
counterfactual stability capture distinct properties. The dominant biological
intervention varies across encoders: shape irregularity produces the largest
response for ResNet-50, immune burden for CTransPath, and nuclear enlargement
for UNI2-h. Tumor--immune mixing produces the smallest biological response
across all three encoders. Rotation generally causes less output movement than
stain brightness, although its nonzero flip rates reveal residual orientation
sensitivity.

Survival effects are attenuated because each intervention replaces only one tile within a fixed 16-tile patient representation. Immune burden nevertheless produces the largest response for every survival model. PathLUPI achieves a lower C-index than the task-trained CTransPath and UNI2-h models but the highest BNR, driven primarily by its response to immune burden. Given that PathLUPI is used without cohort-specific training, this suggests that its released survival representation preferentially captures immune-related variation, although the higher BNR does not compensate for lower predictive discrimination. Patient-clustered 95\% confidence intervals remain above one for every model and task, supporting greater average sensitivity to the evaluated biological interventions than to stain brightness and image rotation.

\section{Conclusion}
\label{sec:conclusion}

CPathOGen provides a structured framework for generating spatially and morphologically controlled H\&E counterfactuals and using them to audit computational pathology models. The fidelity experiments show that requested cellular organization and appearance changes can be recovered across independent analyzers, while downstream probing reveals that predictive performance and intervention sensitivity describe different model properties. This enables safety-oriented evaluation beyond conventional accuracy: models can be examined for sensitivity to biologically motivated factors, robustness to nuisance variation, and potential reliance on spurious signals such as stain or orientation. Such evidence may help identify models requiring additional augmentation or refinement and support model selection based on both performance and feature reliance.

The framework nevertheless requires a large, representative training corpus with sufficiently reliable cellular annotations or weak labels, and its counterfactuals remain condition-matched synthetic samples rather than exact reconstructions. Unmodeled properties may change, and observed prediction shifts characterize model behavior under synthetic interventions rather than patient-level biological causality. Future work should include external-cohort and expert validation, richer cellular conditions, and more efficient candidate selection. More broadly, the same controlled-generation and paired-probing formulation could be adapted to other imaging fields in which spatial organization and global attributes jointly influence model predictions.

{\small
\bibliographystyle{ieeenat_fullname}
\bibliography{main}
}
\end{document}